\documentclass[article]{JHEP3}
\pdfoutput=1
\usepackage{graphicx}
\usepackage{amsmath,epsfig}
\usepackage{amssymb,amsfonts}
\usepackage{latexsym, cite}
\usepackage{epstopdf}  
\usepackage{caption}
\usepackage{subcaption}
\usepackage{placeins}
\usepackage{bm}

\relax
\renewcommand{\theequation}{\arabic{section}.\arabic{equation}}

\def\be{\begin{equation}}
\def\ee{\end{equation}}

\allowdisplaybreaks[2]

\def\bear{\begin{align}}    
\def\eear{\end{align}}
\newcommand{\nn}{\nonumber}

\def\nn{\nonumber}
\newbox\pippobox

\def\II{\relax{\rm I\kern-.18em I}}

\def\e{\epsilon}
\def\l{\lambda}

\def\m{\mu}
\def\n{\nu}

\def\pa{\partial}

\def\a{\alpha}

\def\tr{\ensuremath{\mathrm{Tr}}}

\def\t{\tau}

\def\h{\kappa}
\def\gf{w}

\def\G{G} 

\def\Q{\mathcal{Q}}

\def\mga{\mathcal{G_A}}
\def\mgs{\mathcal{G_S}}

\def\lab{\label}
\def\6{\partial}
\def\la{\langle}
\def\ra{\rangle}
\def\ncs{N_{\textrm{CS}}}
\def\gcs{\Gamma_{\textrm{CS}}}
\def\N{{\mathcal{N}}}

\title{Thermodynamics and CP-odd transport in Holographic QCD with Finite Magnetic Field}

\author{Tara Drwenski$^{1}$, Umut G\"ursoy$^{1}$ and Ioannis Iatrakis$^{2}$
\\
\it $ ^{1}$\textit{Institute for Theoretical Physics, Utrecht University\\ Leuvenlaan 4, 3584 CE Utrecht, The Netherlands}
\\ 
\it $ ^{2}$\textit{Department of Physics and Astronomy, Stony Brook University, Stony Brook, New York 11794-3800, USA}}
{\footnotetext{$^{1}$ T.M.Drwenski@uu.nl, U.Gursoy@uu.nl,  $^{2}$ioannis.iatrakis@stonybrook.edu}}

\preprint{}

\abstract{We consider a bottom-up holographic model of QCD at finite temperature T and magnetic field B, and study dependence of thermodynamics and CP-odd transport on these variables.  As the magnetic field couples to the flavor sector only, one should take the Veneziano limit where the number of flavors and colors are large while their ratio is  kept fixed. We investigate the corresponding holographic background in the approximation where the ratio of flavors to colors is finite but small. We demonstrate that B-dependence of the entropy of QCD is in qualitative agreement with the recent lattice studies. Finally we study the CP-odd transport properties of this system. In particular, we determine the Chern-Simons decay rate at finite B and T, that is an important ingredient in the Chiral Magnetic Effect.}

\begin{document}

\def\g{\gamma}
\def\go{\g_{00}}
\def\gi{\g_{ii}}

\maketitle 

\section{Introduction and Summary}  
\label{section1}

Strongly interacting quantum field theories when coupled to finite magnetic field exhibit a host of interesting phenomena\footnote{We refer to \cite{Kharzeev:2013jha} for an extensive study of the various phenomena mostly for QCD.}. In QCD such effects include the modification of the phase diagram of QCD at finite B \cite{Bali:2011qj}, the chiral magnetic effect \cite{Kharzeev:2007jp, Fukushima:2008xe}, and the magnetic catalysis\cite{Gusynin:1995nb}, or de-catalysis \cite{Bali:2011qj} among others. These problems are much beyond mere academic interest as strong magnetic fields $\vec B$ are produced 
in all non-central heavy ion collisions (i.e.~those with nonzero impact parameter $b$) by the charged ``spectators'' (i.e. the nucleons from the incident nuclei that ``miss'',
flying past each other rather than colliding). 
Indeed,  estimates obtained via 
application of the Biot-Savart law to heavy ion collisions with
$b=4$~fm yield $e|\vec B|/m_\pi^2 \approx$ 1-3 about 0.1-0.2 fm$/c$
after a RHIC collision with $\sqrt{s}=200$~AGeV 
and $e|\vec B|/
m_\pi^2 \approx $ 10-15 at some even earlier time after 
an LHC collision with $\sqrt{s}=2.76$~ATeV~\cite{Kharzeev:2007jp, Skokov:2009qp, Tuchin:2010vs ,Voronyuk:2011jd, Deng:2012pc, Tuchin:2013ie, McLerran:2013hla, Gursoy:2014aka}.

We study the dependence of thermodynamics and the CP-odd transport in QCD in the deconfined phase at finite temperature and magnetic field in the limit of large QCD coupling constant and large number of colors $N_c$. In this limit, the gauge-gravity duality~\cite{Maldacena:1997re, Gubser:1998bc, Witten:1998qj} allows one to study the theory by mapping it to a gravitational theory in one-higher dimension. In particular we employ the bottom-up holographic model put forward in \cite{Gursoy2007part1, Gursoy2007part2, Gursoy2008, Gursoy2009} to model the {\em glue sector} of large-$N_c$ QCD at large 't Hooft coupling.
However, this is not sufficient to study the effects of the magnetic field on the system as B couples the quark-gluon plasma through the quarks that constitute the {\em flavor sector} in the fundamental representation of the gauge group $SU(N_c)$, not the glue sector that is in the adjoint representation.  The dynamics of flavor degrees of freedom  are introduced by considering $N_f$ pairs of flavor branes and anti-branes in the glue background, \cite{Casero:2007ae, Iatrakis:2010jb}. Then, non-trivial dependence on B of any quantity in the large $N_c$ limit will only be visible if we also consider large number of flavors $N_f$ and keep the ratio fixed: 
\be\lab{Ven} 
\N_c\to \infty, \qquad  \N_f\to \infty, \qquad \frac{N_f}{N_c}\equiv x = const\, . 
\ee 
In the gravity dual, this limit requires back reacting the flavor branes on the background solution \cite{Jarvinen2011, Jarvinen2013, Arean:2012mq, Jarvinen2013b, Alho:2013hsa, Jarvinen:2015ofa} and the solution now acquires a non-trivial dependence on the ratio $x$ above. Dependence on any quantity on $B$ will arise in the form $x\, B$. {\em In this paper we consider finite but small} $x$. This approximation, even tough makes our study slightly unrealistic (as $x=1$ for QCD with 3 flavors), simplifies  the calculations drastically as explained below. Therefore, in this work, we shall mostly confine ourselves in the qualitative features of the B-dependence in the system.  Previous holographic studies of the $\mathcal{N}=4$ Super Yang-Mills thermodynamics in the presence of magnetic field include \cite{DHoker2009}. The sphaleron rate in $\mathcal{N}=4$ SYM, for finite $B$ was calculated in \cite{Basar2012}.

\begin{figure}[Ht]
	        \begin{center}
	                \includegraphics[width=0.7\textwidth]{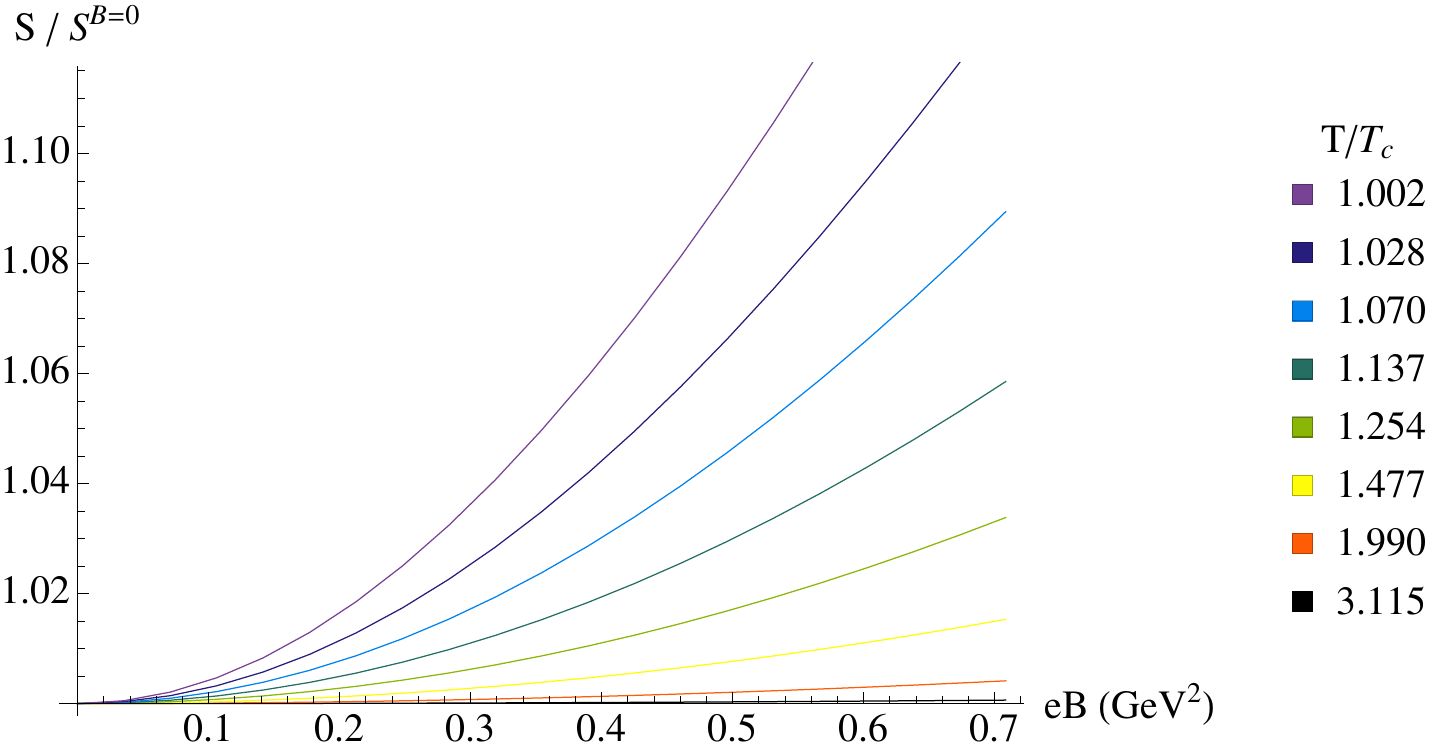}	                
	                \includegraphics[width=.7\textwidth]{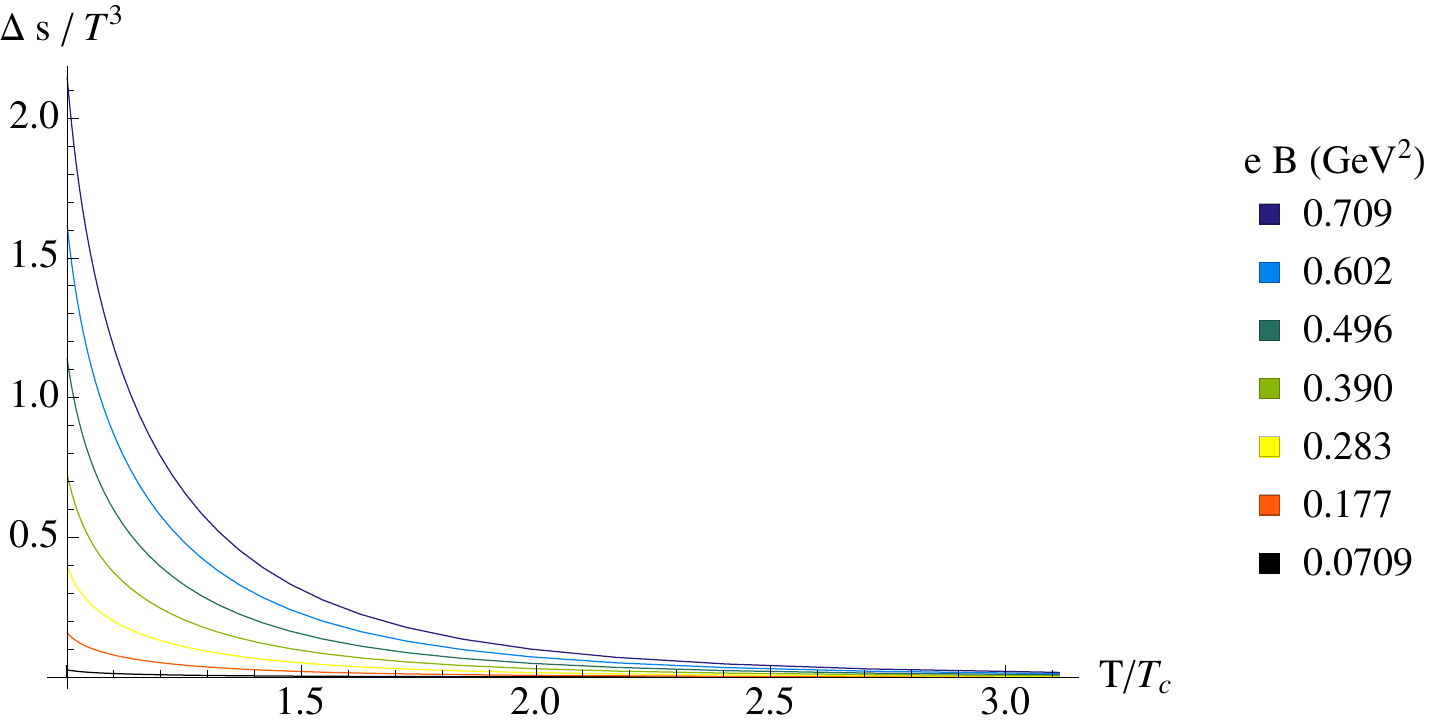}
	        \caption{[Color online] Top: the entropy density $S$, as a function of $eB$ with IHQCD potentials. Down: change in entropy density divided by $T^3$ vs. $T/T_c$ for different values of $eB$.}
	        \label{fig1}
	        \end{center}
	        \end{figure}

\begin{figure}[Ht]
			\centering
	                \includegraphics[width=.6\textwidth]{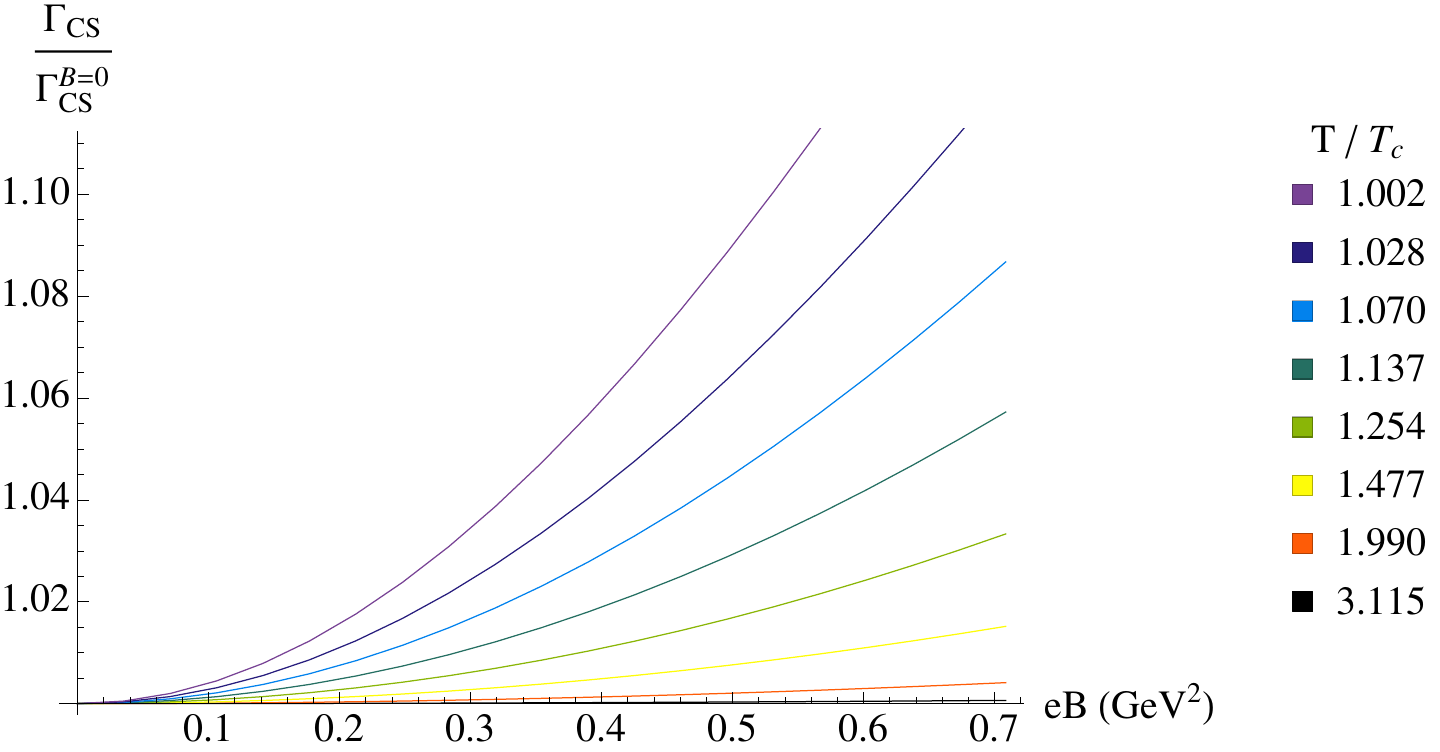}
    				\caption{[Color online] $\Gamma_{CS}$ vs. $eB$ with IHQCD potential for different temperatures. With $c_1 =1$ and $c_4 = 1$. }\label{fig2}
		\end{figure}

We study two separate effects of the magnetic field in this paper. Firstly we ask how the entropy density $S$ of the thermal state depends on B. We find that the ratio $S(B\neq 0)/S(B=0)$ increases with $B$ at any fixed temperature above $T_c$. Here $T_c$ is the deconfinement temperature. We also find that the rate of increase becomes more substantial at lower temperatures $T\gtrsim T_c$ and the dependence of    $S(B\neq 0)/S(B=0)$ on $B$ becomes milder as $T$ is raised. This finding is summarized in figures \ref{fig1}  and \ref{fig2}. All of this is in non-trivial qualitative agreement with the recent lattice studies \cite{Bali:2011qj}. 
Secondly, we consider anomalous transport properties of QCD at finite B and T. In particular we study how the sphaleron decay rate (sometimes called the Chern-Simons decay rate) depends on these variables. 

 In QCD at finite temperature there exist sizable effects, e.g. the sphaleron decays \cite{Moore2010} that generate a non-trivial expectation value $\la \tr\, G\wedge G \ra$, which in turn generates an effective chiral chemical potential $\mu_5$ \cite{Kharzeev:2007jp, Fukushima:2008xe} for the quarks.  This effective thermodynamic variable is an important ingredient in the recently discovered Chiral Magnetic Effect \cite{Kharzeev:2007jp, Fukushima:2008xe}.  In short, the CME refers to generation of a macroscopic electric current in the presence of an external magnetic field $\vec{B}$ in gauge theories with chiral fermions as a result of the chiral anomaly. A number of independent derivations \cite{Fukushima:2008xe, Son:2009tf} reveal that the generated electric current is of her form   
\be\lab{J1} 
\vec{J} = \sigma_\text{CME}\, \vec{B}  \ ,
\ee
where the so-called chiral magnetic conductivity $\sigma_{CME}$ is of the form
\be\lab{sigma1} 
 \sigma_\text{CME} =  \frac{e^2}{2\pi^2} \, \mu_5\, .
\ee 
Therefore, it is essential to determine the dependence of $\mu_5$ on B and T in order to assess the importance of this phenomenon.  On the other hand, the most effective mechanism that generates $\mu_5$ in QCD are sphaleron decays \cite{Moore2010}. Therefore the question translates into a calculation of the sphaleron decay rate in QCD at finite B and T.    This rate in any Quantum Field Theory is captured by the Wightman two point function of the topological charge,
\be
\label{q}
q(x^{\mu})\equiv \frac{1}{16\pi^2}\textrm{tr} \left[F \wedge F\right] = \frac{1}{64\pi^2} \, \epsilon^{\mu\nu\rho\sigma} \textrm{tr} F_{\mu\nu} F_{\rho\sigma},
\ee
where $x^{\mu}=(t,\vec{x})$. In a state invariant under translations in space and time, the rate of change of $\ncs$ per unit volume $V$ per unit time $t$ is called the Sphaleron decay rate, denoted $\gcs$,
\be
\label{gcsdef}
\gcs \equiv \frac{\langle \left(\Delta \ncs\right)^2\rangle}{V t} = \int d^4x \, \left \langle q(x^{\mu}) q(0) \right \rangle_{\textrm{W}},
\ee
where the subscript W denotes the Wightman function. 
Therefore, in order to study the magnitude of the chiral magnetic current, one should study this Wightman correlator of the topological charge. Holographic calculations of $\Gamma_{CS}$ in different holographic models are presented in \cite{Son:2002sd, Craps:2012hd, Gursoy2012}, In the regime of strong interactions, this quantity can be obtained by means of 
the gauge-gravity correspondence by studying the propagation of a bulk axion field in the 5 dimensional gravitational background. 
Our findings are shown in figure \ref{fig2}. We find that, similar to the entropy above, the sphaleron decay rate increases as a function of B and the rate of increase is more substantial at low temperatures. 
Thus, we conclude that presence of the magnetic field in QGP magnifies the chiral magnetic effect, in an indirect fashion, by increasing  $\mu_5$ in (\ref{sigma1}), in addition to the direct dependence on B in (\ref{J1}).

The rest of the paper is organized as follows. In the next section we introduce our holographic model that corresponds to QCD with flavors in the Veneziano limit (\ref{Ven}) and present the equations of the motion for backgrounds at finite T and B. In this section we also introduce the CP-odd part of the gravity action, that contains the axion field, that will be used in calculating the Sphaleron decay rate. In section 3 we present the calculation of the background that we obtain by solving the background equations numerically for a specific choice of the model. In the same section we present our findings for the B and T dependence of thermodynamic variables, in particular they entropy density $S(B,T)$. Section 4 is devoted to the sphaleron decay rate, that we calculate by solving the axion fluctuation equation numerically, on top of the numerical background found in section 3. We present our results for dependence of the sphaleron decay rate on B and T in this section. Section 5 is a discussion and an outlook on our research. Two appendices present details of our calculations. 

\paragraph{Note added:} Another paper \cite{Rougemont:2015oea} that studies the dependence of thermodynamics on magnetic fields appeared on the arXiv yesterday while this work ready to post. While we have some overlap with this paper in our results concerning entropy as a function of B, in general we consider different holographic models  and the focus of the two papers are somewhat different. 

%
%


\section{Holographic QCD with Flavors}

 Improved Holographic QCD, \cite{Gursoy2007part1, Gursoy2007part2, Gursoy2008, Gursoy2009} , is a string theory inspired bottom-up  model of  large $N_c$ 4-dimensional pure Yang-Mills at strong coupling, which is in remarkable agreement with low energy QCD phenomenology. The model was generalized in order to include mesonic physics in the Veneziano limit, where  $N_f \to \infty$, $N_c \to \infty$ and $N_f/N_c = \rm{finite}$, $ \lambda= g^2 N_c = \rm{fixed}$,  \cite{Jarvinen2011, Jarvinen2013, Jarvinen2013b, Alho:2013hsa, Jarvinen:2015ofa} . The full action for the Veneziano QCD model (V-QCD) can be written as

\be
 S = S_g + S_f + S_a
\ee
where $S_g$, $S_f$, and $S_a$ are the actions for the glue, the flavor and the CP-odd sectors respectively. As discussed in \cite{Gursoy2007part1} and \cite{Jarvinen2011}, only the first two terms contribute to the vacuum structure of the theory if the phases of the quark mass matrix and the $\theta$ angle vanish, hence the ground state is CP-even. In this work, we study the theory above the confinement-deconfinement phase transition in the presence of an external magnetic field which descends from the flavor degrees of freedom. However, $S_a$ is important for the analysis of the CP-odd excitations of the model.

\subsection{The glue sector} \label{sec:VQCDglue}

The holographic glue action is a two-derivative gravity-dilaton action with fields that correspond to the lowest dimension operators of the boundary field theory. The metric is dual to the energy momentum tensor of the theory and the dilaton corresponds to the ${\mathbb Tr} F^2$ operator.  The action was introduced in \cite{Gursoy2007part1} and reads
\be
S_g= M^3 N_c^2 \int d^5x \ \sqrt{-g}\left(R-{4\over3}{
(\partial\lambda)^2\over\lambda^2}+V_g(\lambda)\right) \, .
\label{vg}
\ee
Here $\l=e^\phi$ is the exponential of the dilaton field and its boundary value is identified with the holographic 't Hooft coupling. Both the metric and the dilaton field are non-trivial in the ground state\footnote{Our choice for the dilation potential as well as the other potentials in the action is presented in section \ref{choice}.}.

\subsection{The flavor sector} \label{sec:VQCDflavor}

The flavor action is the generalized Sen's action which was first used in holographic QCD in \cite{Casero:2007ae}, in the probe limit, and was incorporated in the study of backreacting flavors in \cite{Jarvinen2011}. The form of the action is 
 
\be
S_f= - \frac{1}{2} M^3 N_c  {\mathbb Tr} \int d^4x\, dr\,
\left(V_f(\l,T^\dagger T)\sqrt{-\det {\bf A}_L}+V_f(\l, TT^\dagger)\sqrt{-\det {\bf A}_R}\right)\,,
\label{generalact}
\ee
The quantities inside the square roots are defined as
\begin{align}
{\bf A}_{L\,MN} &=g_{MN} + \gf(\l,T) F^{(L)}_{MN}
+ {\h(\l,T) \over 2 } \left[(D_M T)^\dagger (D_N T)+
(D_N T)^\dagger (D_M T)\right] \,,\nonumber\\
{\bf A}_{R\,MN} &=g_{MN} + \gf(\l,T) F^{(R)}_{MN}
+ {\h(\l,T) \over 2 } \left[(D_M T) (D_N T)^\dagger+
(D_N T) (D_M T)^\dagger\right] \,,
\label{Senaction}
\end{align}
with the covariant derivative
\be
D_M T = \partial_M T + i  T A_M^L- i A_M^R T\,.
\ee
The gauge fields  $A_{L}$, $A_{R}$, and the complex scalar $T$ transform under the  $U(N_f)_L \times U(N_f)_R$ flavor symmetry and they are dual to the left/right axial current and the quark mass operator, respectively. One can define the vector and axial vector combination of the gauge fields

\be
V_M={A^L_M+A^R_M \over 2} \,\,\,,\,\, A_M={A^L_M-A^R_M \over 2}\,.
\ee
We also define the ratio of the number of flavors over the number of colors
 \be
 x\equiv {N_f\over N_c}\;.
 \ee
The form of the tachyon potential is generally
\be
V_f(\l,TT^\dagger)=V_{f0}(\l) e^{- a(\l) T T^\dagger} \, .
\label{tachpot}
\ee
The Plank mass, M, which appears as an overall factor in front of both $S_g$ and $S_f$, is fixed by requiring the pressure of the system  to approach the large temperature limit of a system free non-interacting fermions and bosons. This fixes $(M \ell)^3 = (1+7 x/4)/ 45 \pi^2$,  \cite{Jarvinen2013}, where $\ell$ is the AdS radius.  The coupling functions $\h(\l,T)$ and $\gf(\l,T)$ are allowed in general to depend on $T$,  through such combinations that the expressions~\eqref{Senaction} transform covariantly under flavor symmetry. However, following string theory intuition we take them independent of $\t$. The potentials $V_{f0}(\l)$, $a(\l)$, $\h(\l)$, and $\gf(\l)$ are constrained by IR properties of QCD like chiral symmetry breaking and meson spectra as it is studied in \cite{Jarvinen2013b}.

\subsection{The CP-odd sector}
\lab{CPodd}

 The CP-odd part of the action has been studied in detail in the probe limit in \cite{Casero:2007ae}. Its main features are that it arises from the Wess-Zumino action that couples the Ramond-Ramond forms with the gauge fields on the branes and it is such that it reproduces the $U(1)_A$ anomaly of the field theory. For finite $x$, the action was introduced in \cite{Jarvinen2013b},
 and it couples the axion field, dual to $\mathbb{T}r F\wedge F$,  from the closed string sector to the tachyon's phase ($T=\t\, e^{i \xi }  \, \mathbb{I}_{N_f}$)  and the $U(1)_A$ axial-vector field, in such a way that reproduces the correct $U(1)_A$ axial anomaly of the boundary field theory 

\be
S_a=-{M^3\,N_c^2\over2}\int d^5x\, \sqrt{g}\,Z(\l)\left[da-x\left(2V_a(\l,T)\,A-\xi \, dV_a(\l,T)
\right)\right]^2 \, ,
\label{samain}
\ee
where $\xi$ is the phase of the tachyon field. The action is normalized so that $a$ is dual to  $\theta/N_c$ with $\theta$ being the standard $\theta$-angle of QCD.  The potential $Z(\lambda)$ has been studied in the $x=0$ case extensively and is constrained by the topological susceptibility of QCD and the $0^{+-}$ glueball spectrum \cite{Gursoy2007part2}. In this work, we are not interested in the exact form of the potential  $V_a$. We notice though that it should satisfy $V_a(\l,T=0)=1$,  in order to have the correct U(1)$_A$ anomaly. A detailed analysis of this action in the zero temperature case is under preparation.

\subsection{Background at finite magnetic field and temperature}

The non-trivial bulk fields on the ground state include the metric, the dilaton, the tachyon and a constant magnetic field. The non-zero background magnetic field arise from the flavor part of the action. Hence, it is naturally defined as the  $U(1)$ part of the vector field

\be
V_M= \left(0, -{x_2 \over 2} B,{x_1 \over 2} B,0,0 \right) \,\, , \,\,\, A_M=0 \, .
\label{magnf}
\ee
 The non-trivial magnetic field breaks the $SO(3)$ rotation symmetry of the vacuum state to $SO(2)$ symmetry of the directions transverse to the magnetic field plane, $x^1-x^2$.  Hence, the Ansatz for the background metric and the dilaton should be,

\be
ds^2= e^{2 A(r)}  \left( {dr^2 \over f(r)} -  f(r) dt^2+dx_1^2 +dx_2^2 +e^{2 W (r)}  dx_3^2\right) \,\, , \,\,\, \lambda=\lambda(r) \,,
\label{bame}
\ee
The UV boundary lies at $r=0$ (and $A\to\infty$). In the UV, the AdS coordinate, $r$, is identified roughly as the inverse of the energy scale of the dual field theory.
At zero temperature, $f(r)=1$ and the solution has $AdS_5$ asymptotics near the boundary with logarithmic corrections. In the IR region of the  space-time the solution asymptotes to a  qualitatively similar solution for $A(r)$ and $\lambda(r)$ as the $B=0$ case, while the function $W(r)$ will be proportional  to $B$. At finite temperature, at least two solutions should exist. One is the similar to the zero-temperature solution but with periodic time coordinate and a black hole solution. At some certain temperature, confinement-deconfinement transition is expected to happen, similarly to $\mathcal{N}=4$ case, \cite{DHoker2009}. Above the deconfinement transition the dominant solution is a black hole metric, so $f(r)$ is non-trivial and satisfies $f(r_h)=0$, where $r=r_h$ is the position of the black hole horizon. The black hole temperature is $T=\beta^{-1}=| f'(r_h)|/4 \pi$.
The boundary asymptotics should match the $T=0$ solution.

Moreover, in the general case of a boundary field theory with $N_f$ light quarks, we consider a background tachyon field  of the form
\be
T=\t(r)  \, \mathbb{I}_{N_f} \, ,
\label{tachans}
\ee
which corresponds to quarks with the same mass. Since, $\t(r)$ is dual to the quark mass operator, its profile signals the chiral symmetry breaking in the boundary field theory. Background solutions with non trivial tachyon which is diverging in the the deep IR of space-time generically correspond to chirally broken state of the field theory. Bulk solutions with identically vanishing $\tau(r)$ signal a chirally symmetric phase of the boundary theory.  V-QCD, at $B=0$, has been studied both at zero and non-zero temperatures and it has been found that the theory exhibits a chiral transition at finite temperature, depending on $x$, above which the chiral symmetry is restored, hence $\tau(r)=0$ and the theory is deconfined. In the present work, we analyze the theory in the chirally symmetric phase and non-trivial $B$. A full study of the model would require to find all the different bulk solutions, with the same near boundary conditions, for each $B$. By comparing their free energies one can decide which is the dominant solution. In the present work, we assume that the magnetic field does not change the phase structure dramatically, so at high temperatures the dominant phase is deconfined and chirally symmetric\footnote{This can be further justified since the condensate in QCD decreases for increasing magnetic field at temperature close to $T_c$, a phenomenon called inverse magnetic catalysis, \cite{Bali:2012zg}.}.
The CP-odd fields are set to zero at the ground state, $a=\xi=A_M=0$. The vacuum action then reads

\begin{align}
S &=M^3 N_c^2 \int d^5x \left[  \sqrt{-g}\left(R-{4\over3}{
(\partial\lambda)^2\over\lambda^2}+V_g(\lambda)\right) \right. \nn \\
&-  \left. x \, V_f(\l,\t) \sqrt{- det (g_{\m\n} + w(\l)\, V_{\m\n} + \kappa(\l,\t)\, \partial_{\m} \t \,\partial_{\n} \t) }
\right]
\label{vacact}
\end{align}
The Einstein equations of motion read

\begin{align}
& R_{\m\n}-{1 \over 2} g_{\m\n} R - \left( {4\over 3} {\partial_{\m} \lambda \pa_{\n} \l \over \l^2} -{2 \over 3} {(\partial \l )^2 \over \l^2} g_{\m\n}  +{1\over 2} g_{\m\n} V_g(\l) \right) 
\nn  \\
&-x {V_f(\l,\t) \over 2} \left(- g_{\m\n} \sqrt{D} +{1 \over \sqrt{D}} {d D \over d g^{\m\n} }  \right)  =0 \, ,
\end{align}
where $D=det(\delta_{\l}^{\m}+w(\l) \, g^{\m\n} \, V_{\n\l} + \kappa(\l , \t)  g^{\m\n} \partial_{\n} \t \, \partial_{\l} \t )$. The full set of equations of motion for non-zero $\tau(r)$ is presented in Appendix \ref{app:eomtau}. We now consider the deconfined and chirally symmetric  phase where the fermion condensate is zero and we have a background magnetic field. This magnetic field comes from the vector field on the flavor branes and is described by Sen's action \eqref{generalact}. In case of small magnetic field the DBI can be expanded to quadratic order to the Maxwell action

\be
\begin{split}
S_f & =- M^3 \, N_c \,  {\mathbb Tr} \, \int d^4 x \, dr  V_f(\l,\t) \sqrt{- g} \sqrt{ \mathrm{det} (\delta^{\mu}_{\nu} +w(\l,\t)^2 g^{\m\rho} V_{\rho \nu} )} \\
 & = - M^3 \, N_c \, N_f \, \int d^4 x \, dr  \, V_f(\l,\t)  \,  \sqrt{-g } \, \left( 1 + {w(\l,\t)^2 \over 4} \, V_{\mu \nu}  \, V^{\m\n} \right) \, .
\end{split}
\ee
The gauge field part of the action reads
\be
\begin{split}
S_B = - M^3 \, N_c \, N_f \, \int d^4 x \, dr  \,  \sqrt{-g } \, {V_b(\l,\t)^2 \over 4} \, V_{\m\n} \, V^{\m\n} \,,
\end{split}
\ee
where $V_b(\l)=V_f(\l)  w(\l)^2$. The potentials are taken independent on the tachyon since it is neglected in the current solution. We notice that for small $x$ the contribution of the magnetic field in the background solution is small. The Einstein equations now read

\begin{align}
& R_{\m\n}-{1 \over 2} g_{\m\n} R - \left( {4\over 3} {\partial_{\m} \lambda \pa_{\n} \l \over \l^2} -{2 \over 3} {(\partial \l )^2 \over \l^2} g_{\m\n}  +{1\over 2} g_{\m\n} V_{eff}(\l) \right) 
\nn  \\
&-x {V_b(\l) \over 2} \left( V_{\m}^{\,\, \rho} V_{\nu \rho} - {g_{\m\n} \over 4} V_{\rho\sigma} V^{\rho\sigma}  \right)  =0 \, ,
\label{EE}
\end{align}
where now the dilaton potential is replaced by $V_{eff}=V_g- x V_{f0}$. The equations of motion for the metric ansatz functions are

\begin{align}
&3 A''(r) + {2\over 3} {\l'^2 \over \l^2} + 3 A'^2 + \left(3 A' -W'\right){ f' \over 2f}  +{3 \, x\, V_b(\l)\,B^2 \, e^{-2 \, A} \over  4\, f(r)} -{ e^{2\, A} \over 2 f(r)} V_{eff}(\l)=0  \, ,\\
&W''+ {W' f'\over f} +W'^2 + 3 A' W'  - {x\, V_b(\l)\,B^2 \, e^{-2 \, A} \over  2 \, f(r)}=0 \, , \\
& f''+(3 A'+  W') \, f'-x\, V_b(\l) \, B^2\, e^{-2 A}  =0 \, .
\label{maxeom}
\end{align}
We can integrate the equation of $f(r)$ and find

\be
f(r)=e^{2 W}\left(1-C_1\int_0^r e^{-3 A- 3 W} dr' \right) \, .
\label{fsol}
\ee
where we considered the boundary conditions, $f(0)=1$ and $W(0)=0$. $C_1$ is determined by requiring regularity on the black hole horizon
\be
C_1={1 \over \int_0^{r_h} e^{-3 A- 3 W} dr' } \, .
\label{C1sol}
\ee
Using \ref{fsol}, this integration constant can be related to the enthalpy  as
\be
S=\frac{C_1}{16 \pi G_5 \,T} \, .
\label{CS1}
\ee
The first order constraint equation reads
\begin{align}
&{2 \over 3}{\l'^2 \over \l^2}- \left(3 \, A' +W' \right) {f'(r) \over 2 \, f(r)}-6\,A'^2 -3\, A' \,W' +{ e^{2\, A} \over 2\, f} V_{eff}(\l)  \nn \\
&-{x \, V_b(\l) \, B^2 \, e^{-2 \, A} \over 4 \, f} =0
\label{maxcons}
\end{align}
The dilaton equation of motion  is not independent but can be derived from combining the above equations

\begin{align}
& {\l''(r) \over \l(r)} -{\l'(r)^2 \over \l(r)^2}+\left( 3 A'(r) +W'(r) +   {f'(r) \over f(r)}   \right) {\l'(r) \over \l(r)} + {3 \over 8} {\l(r) \, e^{2\, A(r)} \over f(r)} \, \partial_{\l} V_{eff}(\l) \nn \\
& - {3 \, B^2 \, e^{ -2 \, A(r)}  \, \l(r) \over 16 \, f(r)} \partial_{\l}V_b(\l)=0\, .
\label{dileom}
\end{align}

The equations of motion enjoy the following scaling symmetries

\begin{itemize}
\item rescaling of $f(r)$, 
$f \to {f \over c_f} \, ,\,\, A \to A-{1\over 2}\log \,c_f \, , \,\, B\to{B \over c_f}$.
\item  rescaling of the AdS coordinate, 
$r \to \Lambda \, r \,\, , \,\,\, A \to A - \log \, \Lambda \,, \,\, B\to {B \over \Lambda^2}$ ,
\item rescaling of the $W(r)$, 
$W \to W +  c_{W}$.
\label{scaling}
\end{itemize}

\subsection{UV asymptotics}
The UV asymptotics of the above equations that we are interested in are AdS with logarithmic corrections \cite{Gursoy2007part1, Gursoy2007part2}. The magnetic field influences the $UV$ expansions of the fields at ${\mathcal O}(r^4)$. In case of a very strong magnetic field and a potentials $V_{eff}\, , \,\, V_b$ of the form

\be
V_{eff}={12 \over \ell^2} (1+ v_1 \l + +v_2 \l^2 + \ldots)\,\, ,\,\,\,
V_{b}(\l)=v_{b\,0}(1+v_{b\, 1} \l + \ldots)
\ee

\begin{align}
& A(r) \simeq \log \left(\frac{\ell }{r}\right)  +\frac{4}{9 \log (\Lambda  r)}-\frac{B^2 x\, v_{b\, 0}}{40 \ell ^2}r^4 \, \log (\Lambda  r)  \,\,,\,\,\,
W(r) \simeq \frac{B^2  x\, v_{b\, 0}  }{8 \ell ^2} r^4 \, \log (\Lambda  r) \, , \nn \\
& \l(r) \simeq -\frac{8}{9  v_1 \log (\Lambda  r)}  -\frac{2 B^2 x\, v_{b\, 0} }{15 v_1 \ell ^2} r^4  \,\, ,\,\,\, 
f(r) \simeq 1+ \left(C_T + \frac{B^2 x\, v_{b\, 0} }{4 \ell ^2}\log (\Lambda  r)\right) r^4 \, ,
\end{align}
where $C_T$ is a temperature dependent coefficient.

\subsection{IR asymptotics}
Requiring regularity on the horizon, the near horizon asymptotics of the system are of the form

\begin{align}
&A=A_h+A'_h \, \e+ \ldots \,\, , \,\,\, W = W_{h}+W'_{h} \, \e + \dots \nn \\ 
&\l=\l_h  +\l'_h \, \e + \ldots \,\, , \,\,\, f= f'_h \e +{f''_h \over 2}\e^2+\ldots  \,,
\end{align}
where $\e$ is the distance from the horizon, $\e=r_h-r$. Expanding the equations of motion, (\ref{maxeom}, \ref{maxcons}, \ref{dileom}) we find the coefficients

\begin{align}
& A'_h=\frac{ V_{eff}\left(\lambda _h\right) e^{2 A_h}}{3 f'_h}  -\frac{e^{-2 A_h} B^2 \, x  V_b( \lambda_h)} {3  f_h} \,\,, \,\,\, 
  W'_h=\frac{e^{-2 A_h} B^2 \, x V_b\left(\lambda _h\right) }{2 f'_h} \nn \\
& \l'_h=-\frac{3 e^{2 A_h}  \lambda_h^2 V_{eff}'(\lambda_h)}{8 f'_h}   +\frac{3 e^{-2 A_h} \lambda_h^2   B^2 \, x  V_b'(\lambda_h ) }{16 f'_h} \,, \nn \\
& f''_h =-e^{2 A_h} V_{eff}(\lambda_h)+ \frac{3}{2} e^{-2 A_h} B^2 \, x   V_b(\lambda_h )\, .
\label{nearhoras}
\end{align}

%
%

\section{Numerical solution}

\subsection{Choice of the model}
\lab{choice}

	In Improved Holographic QCD, the asymptotics of the potential $V_g(\lambda)$\footnotemark are fixed for small $\lambda$ (UV of the field theory) to match the perturbative large-$N_c$ $\beta$-function, which gives the dual field theory asymptotic freedom. In large $\lambda$ (IR of the field theory), the potential must have a form such that the dual field theory is confining and has a linear gapped glueball spectrum \cite{Gursoy2007part1,Gursoy2007part2,Gursoy2008}. In addition, as $r \to 0$, we want metric to approach AdS, $e^A \to \frac{\ell}{r}$ and $\lambda \to -1/\text{log}(r)$ to mimic the perturbative running of large-$N_c$ YM coupling. A form with the correct asymptotics that we  use in our numerical calculations is

	\begin{equation}
		V(\lambda) = \frac{12}{\ell^2} \left( 1 + V_0 \lambda + V_1 \lambda^{4/3} \sqrt{\text{log} \left( 1 + V_2 \lambda^{4/3} + V_3 \lambda^2 \right)} \right) \,.
	\end{equation}
		We can constrain $Z(\lambda)$, the dilaton-dependent normalization of the axion's kinetic term, as done in \cite{Gursoy2012}. We will use the following forms for $Z_2$ with correct asymptotics,
	\begin{align}
		Z(\lambda) &= Z_0 \left( 1 + c_4 \lambda^4 \right)  \\
		Z(\lambda) &= Z_0 \left( 1 + c_1 \lambda + c_4 \lambda^4 \right) ,
	\end{align}
	with coefficients $c_1$ and $c_4$ restricted by lattice calculations of axial glueball masses \cite{Gursoy2012}  as 
	\begin{equation}
		0 \lesssim c_1 \lesssim 5, \quad 0.06 \lesssim c_4 \lesssim 50.
	\end{equation}

The potentials appearing in the flavor action, (\ref{generalact}), can be constrained by looking at their asymptotics and comparing to lattice and perturbative results as done in \cite{Jarvinen2011, Jarvinen2013, Jarvinen2013b}. Their UV asymptotics are chosen to match the perturbative anomalous dimension of the quark mass operator. Their IR asymptotics were fixed by QCD features of the flavor sector as chiral symmetry and the meson spectra, \cite{Jarvinen2013b}.  Here we simply present the form of $V_{f_0}(\l)$ and $w(\l)$, that we use in our calculation
	\begin{align}
		w (\lambda) &= \frac{1}{\left( 1+\frac{3 a_1}{4} \lambda \right)^{4/3}}, \\
		V_{f_0}(\lambda) &= W_0 \left( 1+W_1 \lambda + W_2 \lambda^2 \right),  
	\end{align}
	with coefficients
	\begin{align}
		a_1 &= \frac{115-16 x}{216 \pi^2}, \quad W_0 = 3/11 \\
		W_1 &= \frac{24 + (11-2x) W_0}{27 \pi^2 W_0}, \quad W_2 = \frac{24(857 - 46x) + (4619-1714x+92x^2)W_0}{46656 \pi^4 W_0}.
	\end{align}

\subsection{Numerical technique}
 
In order to define a physical, finite action for a non-compact geometry, we must choose a reference background with the same asymptotics for the metric as well as the dilaton.

We solve Eq.\ (\ref{EE}) using the following numerical method. First, we note that the free integration constants of this solution are $f'_h, \,A_h,\, W_h,\, \phi_h=\log \l_h$ (letting the subscript $h$ denote the function at $r_h$), as all other integration constants are fixed in terms of these by the Einstein's equations and the condition of horizon regularity, (\ref{nearhoras}). The only other free parameters are $r_h$, the location of the horizon, and $B$, the magnetic field strength. We choose the reference background to be the $B=0$ solution. Then, at some UV cutoff $r_c$, for each value of $B \neq 0$, we match the geometry of the solutions at the cutoff by demanding 
	\begin{align}
		\sqrt{f(r_c)} \beta &= \sqrt{f_0(r_c)} \beta_0, \quad &\beta &= \beta_0\\
		e^{2A(r_c)} \text{Vol}_2 &= e^{2A_0(r_c)} \widetilde{\text{Vol}}_2, \quad &\text{Vol}_2 &= \widetilde{\text{Vol}}_2\\
		e^{A(r_c) + W(r_c)} \text{Vol}_1 &= e^{A_0(r_c)} \widetilde{\text{Vol}}_1, \quad &\text{Vol}_1 &= \widetilde{\text{Vol}}_1\\
		\phi(r_c) &= \phi_0(r_c) .
	\end{align}
where the subscript $0$ means $B=0$. The left column contains the requirements for matching the intrinsic Euclidean geometry of the two solutions and the right column contains the extra conditions we add for convenience. For example, it is convenient to see how a solution changes with $B$ while keeping temperature $T=1/\beta=\frac{|f'(r_h)|}{4\pi}$ fixed.

Our numerical method is then a shooting method, that is, we vary the free horizon quantities $A_h,\, W_h,\, \phi_h, \, r_h$ until the constraints at the cutoff are satisfied. Now, we can see how the horizon quantities $A_h,\, W_h,\, \phi_h, \, r_h$ change as a function of $B$. 

\subsection{Solutions} 

We can use the UV matching procedure described above with a choice of $x  = 1/10$ to solve for $A_h$, $W_h$, $\phi_h$, and $r_h$. In the figures below, we present the dependence on the {\em physical} magnetic field $e B_{phys}$ where $e$ is the elementary charge and $B_{phys}$, the physical magnetic field of our dual field theory, is obtained by
	\begin{equation}
		B_{phys} = \frac{B}{\ell^2}, 
	\end{equation}
	where $\ell$ can be found in physical units by taking $\ell \approx \frac{0.00161482}{247 \text{MeV}}$, see \cite{Gursoy2009}.

Once the background solution is found, it is a trivial matter to calculate the entropy of the background using the Bekenstein-Hawking entropy formula 
\be\lab{entBH} 
S = \frac{\text{Area}_h}{4G_5} = \frac{e^{2A(r_h)+W(r_h)}}{4G_5} \, ,
\ee 
where Area$_h$ denotes the area of the horizon and $G_5$ is Newton's constant, related to the 5-dimensional Planck mass by $M^3=1/(16\pi G_5 N_c^2)$. This can be converted into a function of T and B as the location of the horizon $r_h$ can be numerically obtained for varying values of B and T. An important point here is that the dependence on B and T is twofold: in addition to the explicit dependence on $r_h$, hence $T$ and $B$, in (\ref{fig:IHQCD:st}), the metric functions $A$ and $W$ also change as we change $B$ and $T$ resulting in a hidden dependence on these variables. 

\begin{figure}[Ht]
	        \begin{center}
	                \includegraphics[width=0.8\textwidth]{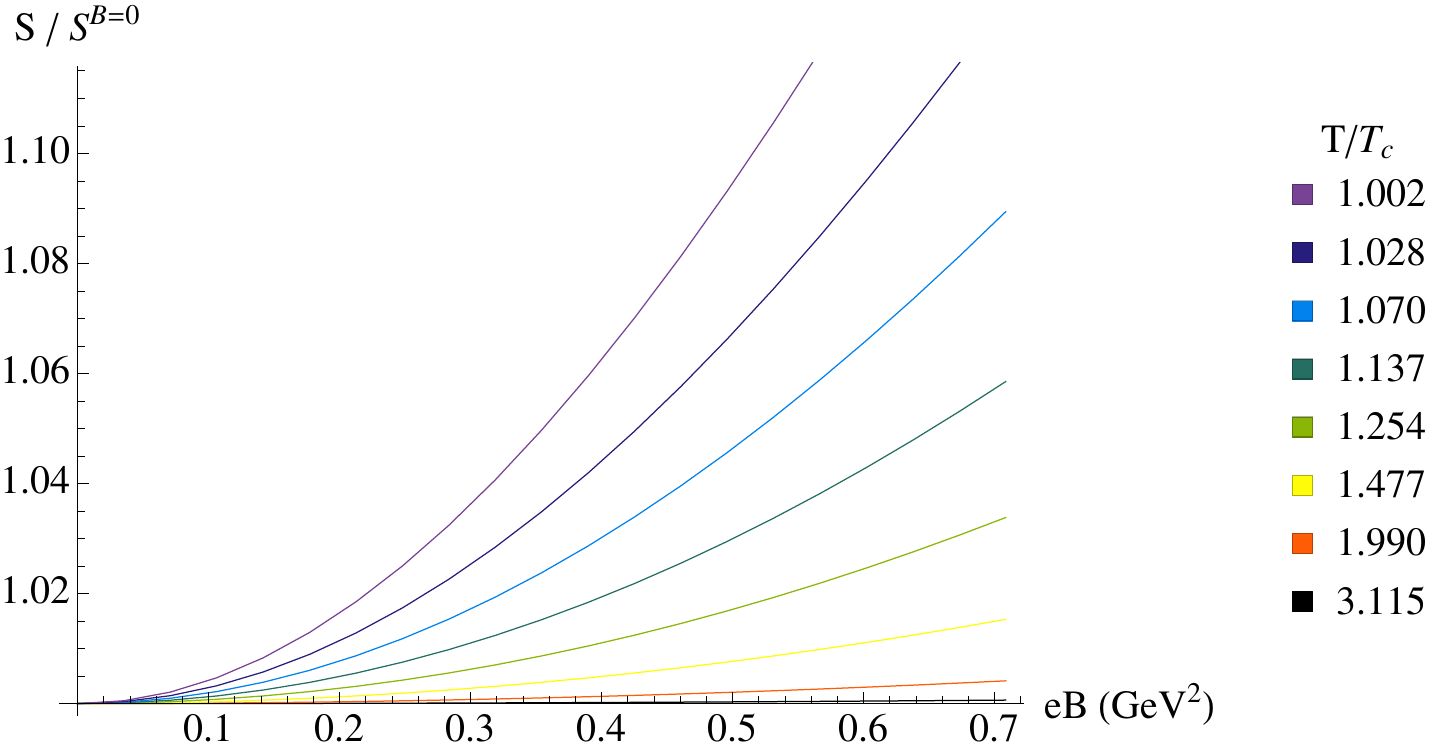}
	        \caption{[Color online] The entropy density $S$, as a function of $eB$ with IHQCD potentials}
	        \label{fig3}
	        \end{center}
	        \end{figure}

\begin{figure}[Ht]
	        \centering
	                \includegraphics[width=.8\textwidth]{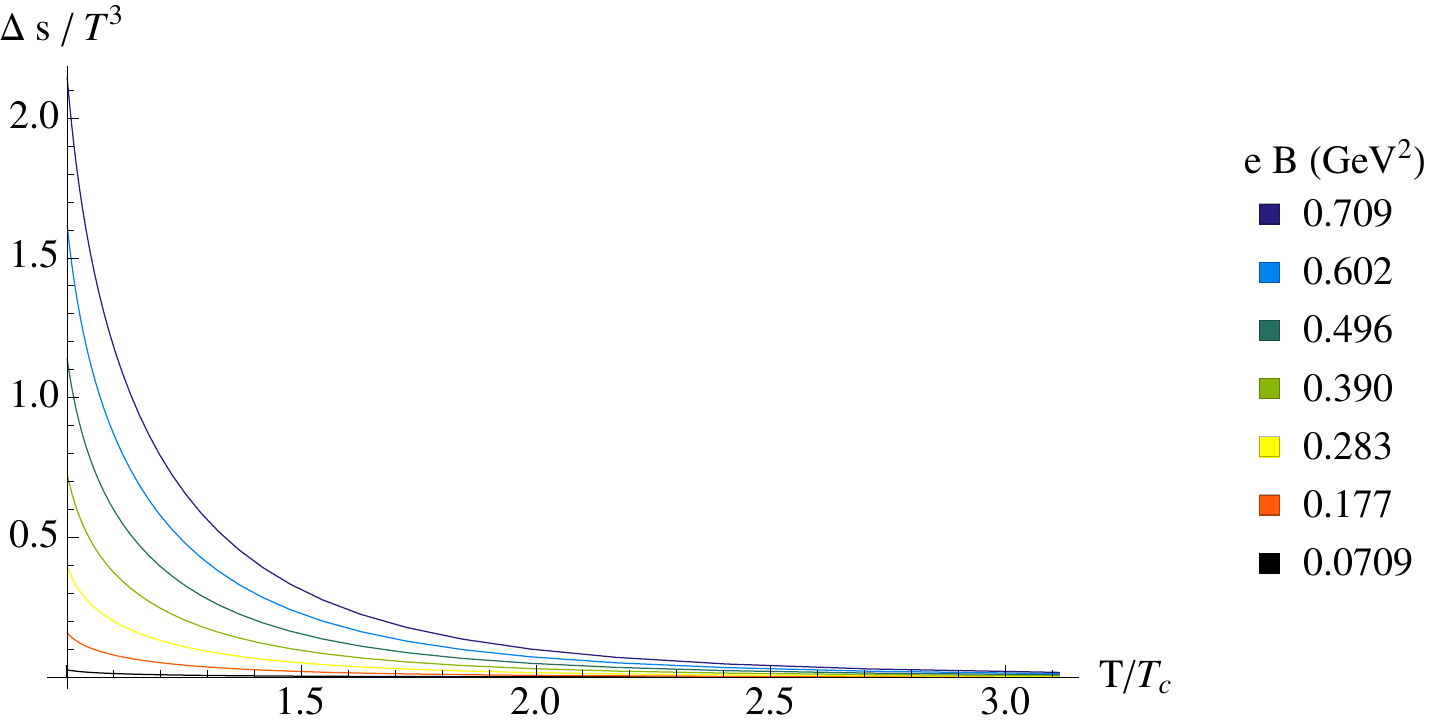}
	        \caption{Change in entropy density divided by $T^3$ vs. $T/T_c$ for different values of $eB$. }\label{fig:IHQCD:st}
		\end{figure}

The plots\footnote{$T_c$ in the figures \ref{fig3} and \ref{fig:IHQCD:st} below refer to the deconfinement temperature at $B=0$. } in Fig.\ \ref{fig3} show the change in entropy as a function of $e B_{phys}$ for different temperatures. 
In addition we plot in Fig.\ \ref{fig:IHQCD:st} the entropy's dependence on temperature, for different values of $eB$. 
 
\begin{figure}[Ht]
	        \centering
	                \includegraphics[width=.8\textwidth]{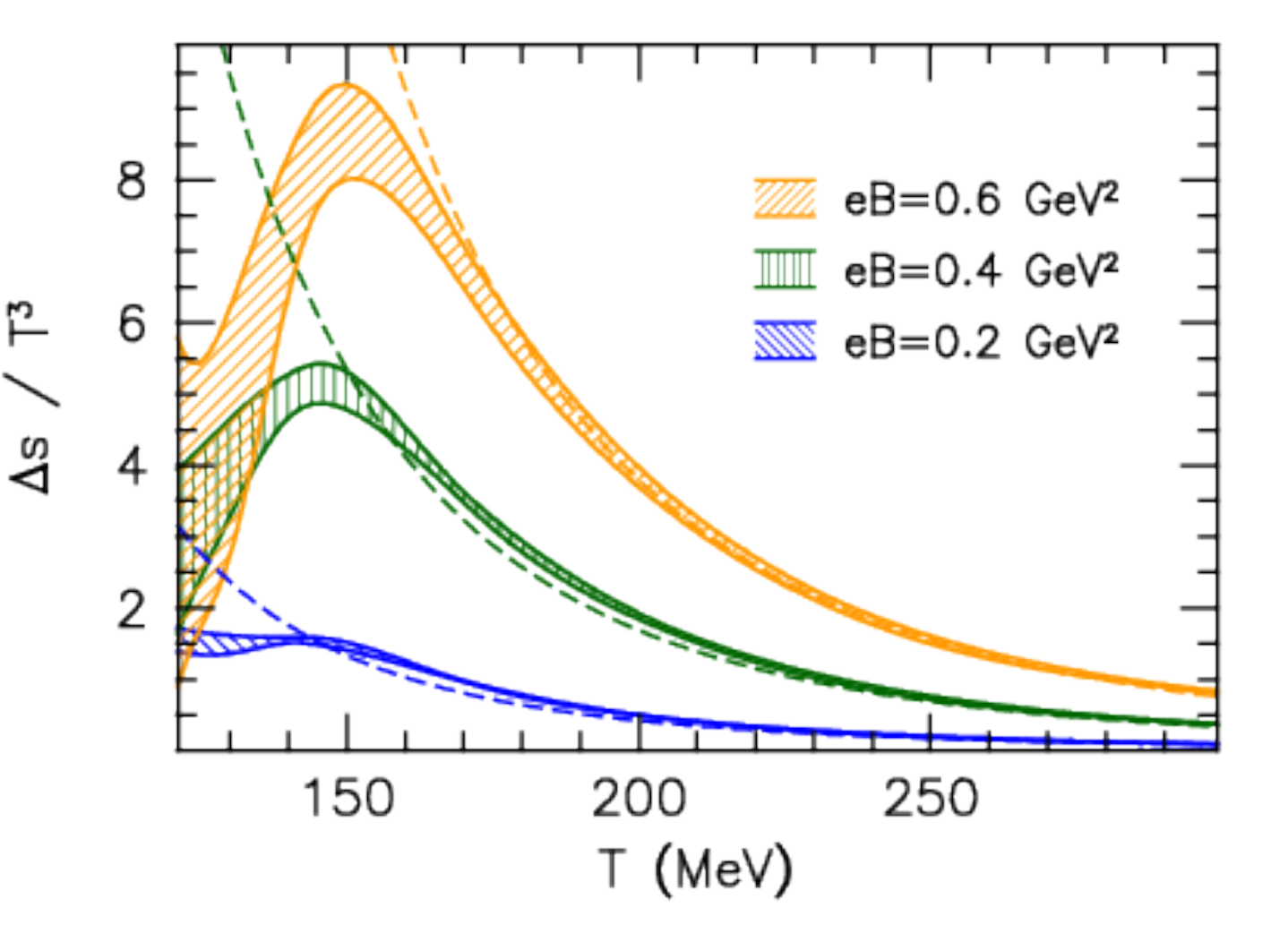}
	        \caption{[Color online] Change in entropy density divided by $T^3$ in terms of $T/T_c$ for different values of $eB$. It is noticed that it has a maximum at $T_c$. The plot was produced using Lattice field theory in \cite{Bali:2014kia}.}\label{fig:latst}
		\end{figure}

It is tempting to compare these results with the recent lattice studies, see for example figure 10 in  
\cite{Bali:2014kia}, that we reproduce here in figure \ref{fig:latst}. Comparison of figures \ref{fig:latst} and \ref{fig:IHQCD:st} show very good qualitative agreement. To see that the agreement is better than it appears in these figures one has to consider only the part $T>T_c$ of figure \ref{fig:latst}. The tails on the left of the peaks at around $T_c\approx 150$ MeV in this figure are absent in the large $N_c$ limit that we consider here.
One can even hope for quantitative agreement and the reason that our results do not quantitatively agree with that of the lattice studies should be because we consider a model where the ratio of flavors to color $x=0.1$, whereas in \cite{Bali:2014kia} this ratio is $x=1$. We expect the agreement become better when we consider larger values of $x$.

\subsection{Free energy and phase transition temperature} 
\label{fe}

Another important thermodynamic quantity is the free energy. In the case where the entropy only depends on the temperature, the free energy can be obtained from the entropy that we found above simply by integrating the first law  in T, \cite{Gursoy2008}. On the other hand, this cannot be done here as $S$ is a function of both B and T. Then the only way to calculate the free energy from the holographic dual theory is from the gravity action (including the Gibbons-Hawking term and the counter-terms) evaluated on the background solution \cite{Gursoy2008}. A similar method is to consider the difference of the on-shell actions evaluated on the {\em black-hole} and the {\em thermal gas} solutions, that is a solution with the same boundary asymptotics but no horizon. As explained in \cite{Witten:1998zw}, (see also \cite{Gursoy2008, Gursoy2009}) the black-hole solution corresponds to the deconfined plasma phase and the thermal gas corresponds to the low temperature, confined phase. The curve in the $(B,T)$ space where this difference vanishes then determines the phase boundary in the system on which the confinement/deconfinement transition takes place. This curve $T_c=T_c(B)$ can be calculated in the aforementioned manner.  

We can easily obtain the thermal gas solution that obeys the same boundary asymptotics as the black-hole solution above by employing the following trick. The thermal gas has no horizon hence vanishing entropy. Using equation (\ref{CS1}) then this means setting the integration constant $C_1$ in the solution (\ref{fsol}) to zero. 
Hence the thermal gas solution obeys 
\be\lab{tg}
f(r) = \exp(2W(r))\, .
\ee
The other metric functions can then be determined by numerical integration. Once both the black-hole and the thermal gas solutions are obtained, one should calculate the difference of the on-shell actions to determine the phase diagram. We shall not carry out this calculation in this paper for two reasons: firstly there are various numerical difficulties which render the calculation very tricky and one needs new techniques to maintain numerical efficiency of the difference of the actions. Secondly, we consider small values of $x$ in this paper, hence the difference $T_c(B) - T_c(0) = {\cal O}(x)$ that is small. This will make a difference  ${\cal O}(x^2)$ in the figures \ref{fig3} and \ref{fig:IHQCD:st} above, which we can safely neglect.

\section{CP-odd fluctuations at finite $x$ and magnetic field $B$.}

We now analyze the coupling of CP-odd excitations of the model and calculate the Chern-Simons diffusion rate at finite $B$. 
As explained in section \ref{CPodd}, the pseudo scalar axion $\alpha$ is dual  to the topological charge operator given by
	\begin{equation}
		q(x^\mu) \equiv \frac{1}{32 \pi^2} \mathbb{T}r \left( F_{\mu\nu} \tilde{F}^{\mu\nu} \right).
	\end{equation}
	Recall that the Chern-Simons diffusion rate is given by
	\begin{equation}
		\Gamma_{CS} \equiv \lim_{V \to \infty} \lim_{t \to \infty} \frac{\langle \left( \Delta N_{CS}\right)^2 \rangle}{V t}.
	\end{equation}
	This can be rewritten as
	\begin{equation}
		\Gamma_{CS} = \int d^4x \, \langle q(x^\mu) q(0) \rangle_W,
	\end{equation}
	where  $\langle \, \rangle_W$ denotes the Wightman correlator, \cite{Moore2010}. 

	In Fourier space, let $\hat{G}_W(\omega, \vec{k})$ and $\hat{G}_R(\omega, \vec{k})$ denote the Wightman and retarded Green's functions of the topological charge operator $q$, defined in Minkowski spacetime as
	\begin{align}
		\hat{G}_R(k) &= -i \int d^4x \, e^{-ik \cdot x} \theta(t) \langle [q(x),q(0)] \rangle \\
		\hat{G}_W(k) &= \frac{1}{2} \int d^4x \, e^{-ik \cdot x}  \langle q(x)q(0)-q(0)q(x) \rangle 
	\end{align}
	where $k=(\omega,\vec{k})$. Then, the fluctuation-dissipation theorem relates these as
	\begin{equation}
		\hat{G}_W(\omega, \vec{k}) = -\text{coth} \left( \frac{\omega}{2T} \right) \text{Im} \,\hat{G}_R(\omega, \vec{k}).
	\end{equation}
	for $\omega \ll T$. Taking the zero momentum and small frequency limit of this, we can write the Chern-Simons diffusion rate as
	\begin{equation}
		\Gamma_{CS} =-\kappa^2 \lim_{\omega \to 0}\, \frac{2T}{\omega} \, \text{Im}\, \hat{G}_R (\omega, \vec{k}=0),
	\end{equation}
	where we now consider $\hat{G}_R (\omega, \vec{k})$ to be the retarded Green's function of the topological charge operator $q(x^\mu)$, dual to the axion. To compute the above transport coefficient in our model at finite B, we study the quadratic excitations of the axion field which are coupled to the $U(1)_A$ axial-vector current and the phase of the tachyon. 
The interplay of the gluon topological correlator and the axial current  has been studied in the probe limit in \cite{Gursoy:2014ela, Jimenez-Alba:2014iia, Iatrakis:2014dka}.
In the DBI, (\ref{generalact}), the coupling of the axial-vector to the phase of the tachyon  is proportional to the tachyon background solution which is taken zero, hence this is neglected. At $\tau(r)=0$, the potential appearing in the CP-odd action,  (\ref{samain}), is $V_a(\l, \tau=0)=1$, so the axion coupling to the phase of the tachyon is also zero since it is proportional to the derivative of $V_a$. 
Expanding the actions (\ref{generalact}, \ref{samain}) up to second order in the fluctuations we have,	

 \begin{align}
S_1 & = {1 \over 4}M^3  N_c^2\, x\, \int d^4 x \, dr  V_b(\l)\, \sqrt{-\mathrm{det}({\mathcal G})} 
 \biggl[ \mgs^{MS} \, A_{ST} \, \mgs^{TN} \, A_{NM}  \\
& + \mga^{MS} \, A_{ST} \, \mga^{TN} A_{NM}  -  {1 \over 2}\, (\mga^{MN} A_{MN})^2 \biggr]  \nn \\
 S_2 & = -{1 \over 2} M^3 N_c^2 \, \int d^4 x dr \sqrt{-g}  \, Z(\l) \, \big(\partial_M \a - 2 x \,A_M  \big) \big(\partial_N \a - 2 x V_a \,A_N \big) g^{MN} \nn \, .
 \end{align}
 $S_1$ comes from the expansion of the $DBI$ and the metric $\mathcal{G}$ is defined as
\be
{\mathcal G}_{MN}= 
\left( \begin{array}{ccccc}
-e^{2 A(r)} f(r) & 0 & 0 & 0 & 0 \\
0 & e^{2 A(r)} & w(\l,\t) B & 0 & 0 \\
0 & -w(\l,\t) B & e^{2 A(r)} & 0 & 0 \\
0 & 0 & 0 & e^{2 A(r)+2 W(r)} & 0\\
0 & 0 & 0 & 0 & {e^{2 A(r)} \over f(r)} \, ,
\end{array}
\right)
\ee
%
%
It is noticed that  ${\mathcal G}_{MN}$ is split in its symmetric and antisymmetric parts ${\mathcal G}_{MN}=\mgs_{MN}+\mga_{MN}$.
The equations of motion read to leading order in the magnetic field

\begin{eqnarray}
&&  \pa_M \biggl[ Z(\l) \, \sqrt{-g} \, g^{MN} (\partial_M \a(r,x) - 2 x  \,A_M(r,x) ) \biggr] =0 
\label{axeom} \\
&& \partial_N \biggl[ V_b(\l) \, \sqrt{-\mathrm{det}({\mathcal G})}  \biggl( \mgs^{MS} \mgs^{TN} A_{ST} + \mga^{MS} \mga^{TN} A_{ST} \nn\\
&&  + {1\over 2}\, \mga^{ST} \mga^{MN} A_{ST} \biggr) \biggr] 
+2 Z(\l) \, \sqrt{-g} \, g^{MN} \biggl[\partial_N \a(r,x) -2 x V_a \,A_N(r,x) \biggr]=0 \nn
\end{eqnarray}
In order to calculate the Chern-Simons diffusion rate we now look at the equation of motion for the axion fluctuation $\alpha$ in the small $\omega$ limit.	Following \cite{Son:2002sd}, we solve this equation with a Dirichlet boundary condition at the asymptotically AdS boundary (at $r=0$) and an in-going wave boundary condition at the horizon $r_h$. Considering the case where $k_{\mu}=(-\omega,0,0,0)$, the solution  for the axion takes the form
	\begin{equation}
		\alpha (r, t) = \int \frac{d \omega}{2 \pi} \, e^{- i \omega t}\, \alpha (r, \omega)\, a(\omega),
	\end{equation}
	where $a(\omega)$ is fixed by the Dirichlet boundary condition,
	\begin{equation}
		\lim_{r \to 0}  \alpha (r, t) = \int \frac{d \omega}{2 \pi}\, e^{-i \omega t}\, a(\omega),
	\end{equation}
	and $\alpha(r,\omega)$ is the solution to Eq.\ (\ref{axeom}) in momentum space. Similar expansions hold for $A_M$. Taking the spatial components of the axial-vector field to be zero, the fluctuation equations become

\begin{align}
& \pa_r \biggl[ Z(\l)\, e^{3 A+W} \,f \, \partial_r \a  \biggr] +Z(\l) {e^{3 A+W} \over f} (\omega^2 \a- 2i\, \omega \,  x\,  A_t)=0 \nn \\
& \pa_r \biggl[ i \omega \, V_b(\l) \, e^{A +W}  \, \pa_r A_t   \biggr]  +2 Z(\l) \,  {e^{3 A+W} \over f} \, ( \omega^2 \a -2i \omega \,x \,A_t)=0 \nn \\
&  i \omega \,V_b(\l) \, e^{A +W}  \, \pa_r A_t -2 Z(\l)\,e^{3 A+W} \,f \, \partial_r \a =0 \, .
\label{flucteom}
\end{align}
In the present work, we restrict ourselves in the calculation of the Chern-Simons diffusion rate, which is the transport coefficient of $q(x)$, hence we are interested in solving the

\begin{align}
\pa_r \biggl[ Z(\l)\, e^{3 A+W} \,f \, \partial_r \a  \biggr] +Z(\l) {e^{3 A+W} \over f}\omega^2 \a=0 \, .
\label{axionEOMkspace}
\end{align}
 The on-shell action of the axion reduces to the boundary term
	\begin{equation}
		S^{\text{on-shell}}_{\text{axion}} =\left. \int \frac{d\omega}{2\pi} \, a(-\omega) \, \mathcal{F}(r, \omega) \, a(\omega)  \right|^{r_h}_0,
	\end{equation}
	where
	\begin{equation}
		\mathcal{F}(r,\omega) \equiv -\frac{M^3}{2}  \alpha(r,-\omega) \, Z_(\lambda(r)) \, \sqrt{-g} \, g^{rr} \, \partial_r  \alpha(r,\omega).
		\label{eq:F}
	\end{equation}
	The retarded Green's function as prescribed by \cite{Son:2002sd}, is given by
	\begin{equation}
		\hat{G}_R(\omega) = -2 \lim_{r \to 0} \mathcal{F}(r, \omega).
		\label{eq:GRandF}
	\end{equation}

	In order to calculate $\Gamma_{CS}$, we need to solve Eq.\ (\ref{axionEOMkspace}) for $\alpha$ with small $\omega$. We do this using near-horizon matching following \cite{Gursoy2012}. In this method, we first solve Eq.\ (\ref{axionEOMkspace}) with $\omega=0$ and expand the solution near the horizon. Then we reverse the order, solving in the near-horizon region and then expanding in small $\omega$. Finally, we match these to solutions to obtain $\mathcal{F}$.

	When $\omega=0$, the solution of Eq.\ (\ref{axionEOMkspace}) is
	\begin{equation}
		\delta \alpha = C_1 + C_2 \int_0^r \frac{dr'}{Z(\lambda(r')) \, e^{3A(r')+W(r')} \, f(r')},
		\label{eq:kwiszero}
	\end{equation}
	with $C_1$ and $C_2$ constants. The integral term in Eq.\ (\ref{eq:kwiszero}) diverges as $r \to r_h$, since $f(r_h) = 0$. Therefore, a normalizable solution must have $C_2 =0$ when $\omega = 0$ and $C_2 \propto \omega$ when $\omega$ is small. This solution Eq.\ (\ref{eq:kwiszero}) gives for Eq.\ (\ref{eq:F}),
	\begin{equation}
		\lim_{r \to 0} \mathcal{F}(r, \omega) =- \frac{M^3}{2} C_1 C_2, \quad \omega \ll T, \, 
		\label{eq:FC1C2}
	\end{equation}
	In order to have unit normalization at the asymptotically AdS boundary, we choose $C_1 =1$.
	Now, expanding Eq.\ (\ref{eq:kwiszero}) near the horizon, we find the solution
	\begin{equation}
		\delta \alpha = C_1 + \frac{C_2}{Z(\lambda_h) \, e^{3A(r_h)+W(r_h)} \, f'(r_h)}\text{log}(r_h - r) + \mathcal{O}(r_h - r),
		\label{eq:311}
	\end{equation}
	where we used $|f'(r_h)| = 4 \pi T$.
	Next we do the reversed order of operations. First, we expand Eq.\ (\ref{axionEOMkspace}) near the horizon and we find the solution
	\begin{equation}
		\alpha = C_{+}(r_h - r)^{\frac{i\omega}{4\pi T}}+C_{-}(r_h - r)^{-\frac{i\omega}{4\pi T}},
	\end{equation}
	with coefficients $C_+$, $C_-$ depending on $\omega$ but not $r$. In order to have an in-going wave condition at the horizon, we set $C_+=0$. Expanding the solution for small $\omega$, we get
	\begin{equation}
		\alpha = C_- - i \frac{\omega}{4\pi T}C_- \text{log}(r_h - r) + \mathcal{O}\left( \frac{\omega^2}{T^2} \right).
		\label{eq:313}
	\end{equation}

	By matching the constant and logarithm terms in Eqs.\ (\ref{eq:311}) and (\ref{eq:313}), we find
	\begin{equation}
		C_1=C_-=1, \quad C_2=-i\omega Z(\lambda_h) e^{3A(r_h)+W(r_h)} C_-.
	\end{equation}
	Using Eqs.\ (\ref{eq:GRandF}) and (\ref{eq:FC1C2}), we obtain
	\begin{equation}
		\hat{G}_R (\omega, \vec{k}=0) = -i\omega M_p^3 Z_2(\lambda_h)e^{3A(r_h)+W(r_h)}, \quad \omega \ll T.
	\end{equation}
	And so our final result for $\Gamma_{CS}$ is
	\begin{equation}
		\label{CSDiffusion}
	\Gamma_{CS}= \frac{1}{N_c^2}\frac{T}{8\pi G_5}\kappa^2 Z(\lambda_h)e^{3A(r_h) + W(r_h)} = \frac{1}{N_c^2}\frac{s T}{2\pi}\kappa^2 Z(\lambda_h),
	\end{equation}
	using that the entropy density is $s=\frac{e^{3A(r_h) + W(r_h)}} {4 G_5}$ and $M^3=1/(16\pi G_5 N_c^2)$.

		\begin{figure}[Ht]
			\centering
	                \includegraphics[width=.8\textwidth]{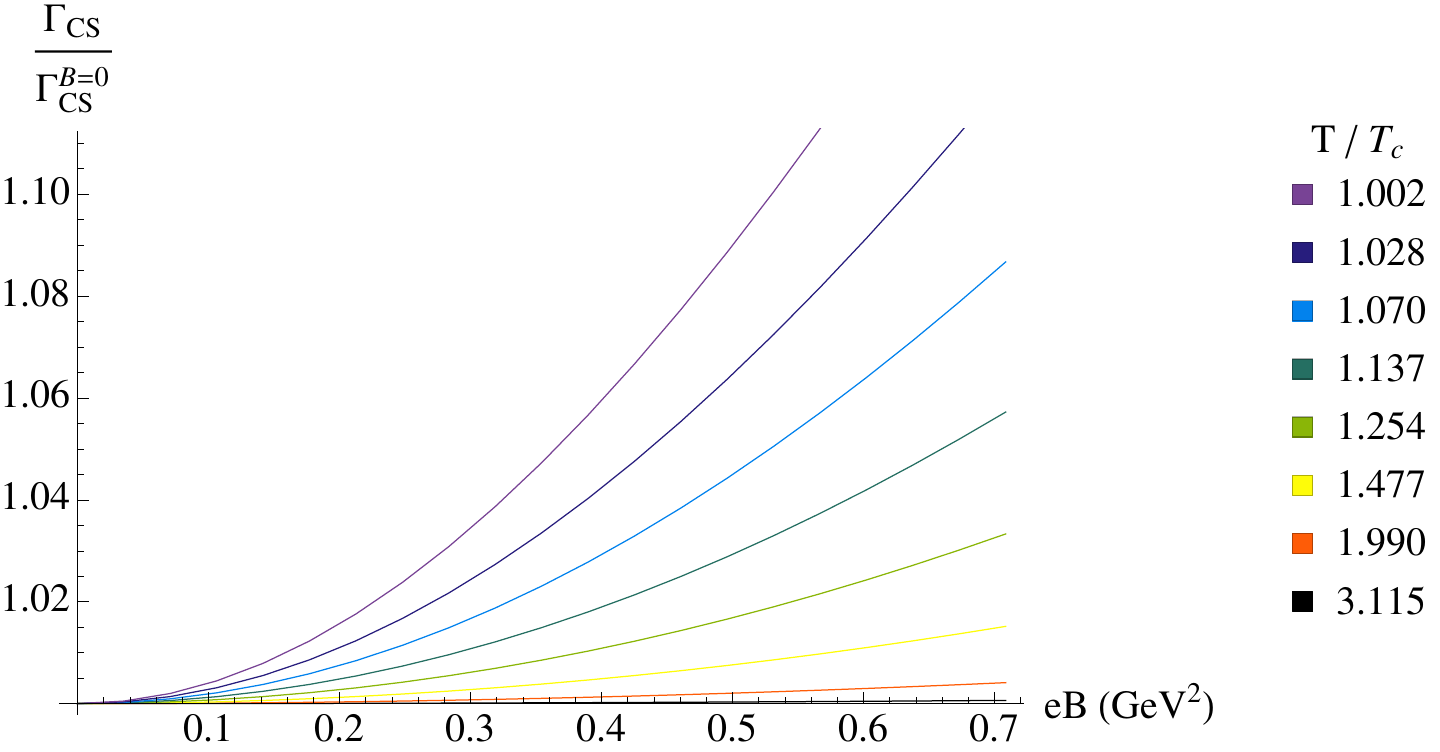}
    				\caption{[Color online] $\Gamma_{CS}$ vs. $eB$ with IHQCD potential for different temperatures. With $c_1 =1$ and $c_4 = 1$. }\label{fig:IHQCD:CS-1-1}
		\end{figure}

The dependence of $\Gamma_{CS}$ on temperature in Fig.\ \ref{fig:IHQCD:CS-1-1} is in good agreement with the previous result in \cite{Gursoy2012}, showing a peak at the devonfinement transition. We notice that $\Gamma_{CS}$ is quadratic in $B$ for weak magnetic field and becomes linear for strong magnetic field.

\section{Discussion and Outlook}

In this paper we studied  the effect of finite magnetic field on the quark-gluon plasma in the Veneziano limit (\ref{Ven}) at strong coupling, using a bottom-up holographic model for QCD. The Veneziano limit is crucial because the magnetic field couples to the plasma only through the flavor sector and in the large color limit, this coupling effectively vanishes unless one takes the number of flavors to infinity as well, keeping the ratio fixed. 
We found that the thermodynamic variables exhibit non-trivial dependence on B. In particular the entropy density increases with B and the rate of increase is more pronounced at smaller values of T, close to the deconfinement temperature $T_c$. This result is in very good qualitative agreement with the recent lattice studies \cite{Bali:2014kia}. We also studied the sphaleron decay rate and found that this rate also increases with B. 
This  means that sphaleron decay processes are favored in the presence of magnetic field which in turn imply a larger value for the effective axial chemical potential $\mu_5$ in (\ref{sigma1}). Therefore the chiral magnetic effect is more probable in a strongly interacting plasma with magnetic field.   

There are several future directions to explore. First of all, we simplified our calculations by taking a small value of $x$, the ratio of flavors to colors. This allowed us to expand the DBI action on the flavor branes. It is straightforward but cumbersome to lift this approximation and study the system in a more realistic case of $x\sim 1$. We leave this study for the future. 

Secondly, we have not studied the dependence of the free energy on B in this paper. As explained at the end of section \ref{fe}, this, and the phase diagram of the theory can be explored by studying the difference of the on-shell actions evaluated on the black-hole (deconfined phase) and the thermal gas (confined phase) solutions. We plan to study this in the near future. In fact, one can also add the baryon chemical potential $\mu$ to the phase space and explore the full phase diagram in the space $(T,B,\mu)$. 

Finally, we are currently studying the response of the axial charge to the fluctuations of the topological operator, which requires the study of the full, coupled system of the CP-odd excitations. Similar issues are addressed in the probe limit in \cite{Gursoy:2014ela, Jimenez-Alba:2014iia, Iatrakis:2014dka}.

\section*{Acknowledgements}

The authors would like to thank  D.~Kharzeev, S.~ Lin and Y.~ Yin for useful
discussions.
This work is supported in part by the DOE grant No. DE-FG-88ER40388
(I.I.).
This work is part of the D-ITP consortium, a program of the Netherlands Organisation for Scientific Research (NWO) that is funded by the Dutch Ministry of Education, Culture and Science (OCW).

\appendix
\renewcommand{\theequation}{\thesection.\arabic{equation}}
\addcontentsline{toc}{section}{Appendices}
\section*{APPENDIX}

\section{Equations of motion for $\tau \not=0$}
\label{app:eomtau}

Einstein equations are written in terms of the ansatz functions of Eqs. (\ref{magnf}, \ref{bame}, \ref{tachans})

\begin{align}
&3 A''(r) + {2\over 3} {\l'^2 \over \l^2} + 3 A'^2 + \left(3 A' -W'\right){ f' \over 2f}  +{x\, V_f(\l,\t) \, G(r) \, e^{2 \, A} \over  2 \, Q(r)\, f(r)}(2 \, Q^2-1) -{ e^{2\, A} \over 2 f(r)} V_g(\l)=0 \nn \, , \\
& \nn \\
&W''- {W' f'\over f} +W'^2 + 3 A' W'  + {x\, V_f(\l,\t) \,G(r) \, e^{2 \, A} \over  2 \, Q(r)\, f(r)} \left( 1-Q^2\right)=0 \, , \\
& \nn \\
& f''+(3 A'+  W') \, f'-{x\, V_f(\l,\t) \, e^{2 A}\, G \over Q}\left(1-Q^2 \right)  =0 \, ,
\label{geneom}
\end{align}
where we have defined  the functions
\be \label{Gdef}
 \G(r) = \sqrt{1 + e^{-2A(r)}\h(\l,\tau) f(r) (\partial_r \t(r))^2} \, \, , \,\,\, \Q(r)=\sqrt{1+w(\l,\t)^2 B^2 e^{-4 A(r)}}.
\ee
The first order constraint equation reads
\begin{align}
&{2 \over 3}{\l'^2 \over \l^2}- \left(3 \, A' +W' \right) {f'(r) \over 2 \, f(r)}-6\,A'^2 -3\, A' \,W' +{ e^{2\, A} \over 2\, f} V_g(\l)  \nn \\
&-{x \, V_f(\l,\t) \, Q \, e^{2 \, A} \over 2 \, G \, f} =0\, .
\label{cons1}
\end{align}
The dilaton eom is

\bear
& {\l''(r) \over \l(r)} -{\l'(r)^2 \over \l(r)^2}+\left(  3 A'(r) +W'(r) +   {f'(r) \over f(r)}   \right) {\l'(r) \over \l(r)} + {3 \over 8} {\l(r) \, e^{2\, A(r)} \over f(r)} \, \partial_{\l} V_g(\l) \nn \\
& - {3\, x \, B^2 \, e^{ -2 \, A(r)}  \, G(r) \, \l(r) \, V_f(\l,\t) w(\l,\t) \over 8 \, f(r) \, Q(r)}\partial_{\l}w(\l,\t)
-{3 \, x \, e^{2 \, A(r)}  \, G(r) \, \l(r) \, Q(r) \over 8 f(r) } \partial_{\l} V_f(\l,\t) \nn \\
&-{3 \, x \, \l(r) \, Q(r) \, V_f(\l,\t) \, \t'(r)^2 \over 16 \, G(r) } \partial_{\l} \kappa(\l,\t)=0 \,.
\end{align}
This is not independent but can be derived from combing Eqs. (\ref{geneom}, \ref{cons1})
The tachyon equation of motion is

\begin{align}
&\t''(r)-{e^{2 \, A(r)} \, G(r)^2 \over f(r)\, \kappa(\l,\t)}{\partial_{\t} \log \, V_f (\l,\t)} +\, e^{-2 \, A(r)} \, f(r) \, \h(\l,\t) \, \left( W'(r) +{1 \over 2} \,{ f'(r) \over f(r)} \right. \nn \\ 
&\left. + 2 A'(r) {1+Q(r)^2 \over Q(r)^2} +{1 \over 2}\, \l'(r) \, \partial_{\l} \log \,( \kappa(\l,\t)  \,V_f(\l,\t)^2 )- {\l'(r) \,(1-Q(r)^2)  \over Q(r)^2} \partial_{\l} \, \log \, w(\l,\t) \right) \t'(r)^3 \nn \\
&+ \left(A'(r){2+Q(r)^2 \over Q(r)^2}+W'(r) +{f'(r) \over f(r)} + \l'(r) \partial_{\l} \log(V_f(\l,\t) \, \h(\l,\t))- {\l'(r) \,(1-Q(r)^2)  \over Q(r)^2} \partial_{\l} \, \log \, w(\l,\t) \right) \,\t'(r) \nn \\
&+{\t'(r)^2 \over 2} \partial_{\t} \log \, \kappa(\l,\t)+ {e^{2 \, A(r)} \, G(r)^2 \, (1- Q(r)^2) \over f(r) \, \kappa(\l,\t) \, Q(r)^2} \, \partial_{\t} \, \log w(\l,\t)=0
\end{align}
To find the equation of motion of the gauge field, we define the matrix in the square root of the flavor action, Eq.(\ref{vacact}), as

\be
{\bf A}_{L\,MN} = {\bf A}_{R\,MN} = {\mathcal G}_{MN}= 
\left( \begin{array}{ccccc}
-e^{2 A(r)} f(r) & 0 & 0 & 0 & 0 \\
0 & e^{2 A(r)} & w(\l,\t) B & 0 & 0 \\
0 & -w(\l,\t) B & e^{2 A(r)} & 0 & 0 \\
0 & 0 & 0 & e^{2 A(r)+2 W(r)} & 0\\
0 & 0 & 0 & 0 & {e^{2 A(r)} \over f(r)} G(r)^2 \, .
\end{array}
\right)
\label{DBImet}
\ee
The inverse matrix is 

\be
{\mathcal G}^{MN}= 
\left( \begin{array}{ccccc}
-{e^{-2 A(r)} \over f(r)} & 0 & 0 & 0 & 0 \\
0 & {e^{-2 A(r)} \over \Q(r)^2} & -{w(\l,\t) B e^{-4 A(r)} \over \Q(r)^2} & 0 & 0 \\
0 & {w(\l,\t) B e^{-4 A(r)} \over \Q(r)^2} & {e^{-2 A(r)} \over \Q(r)^2} & 0 & 0 \\
0 & 0 & 0 & e^{-2 A(r)-2 W(r)} & 0\\
0 & 0 & 0 & 0 & {e^{-2 A(r)}  f(r)} \over G(r)^2 \, ,
\end{array}
\right)\, , 
\ee
Then, the eom of the magnetic field is

\be
\partial_{M} \left( V_f(\l,\t) \, w(\l,\t) \, \sqrt{- \mathcal{G} } \,  \mathcal{G}_A^{MN}  \right)=0 \, ,
\ee 
where $\mathcal{G}_A$ is the antisymmetric part of $\mathcal{G}$. The above equation is automatically satisfied for the gauge field of the form \eqref{magnf}.

\section{Fluctuation equations $\tau \not=0$ and $B\not=0$}
\label{app:}

The quadratic action of those fluctuations reads

 \begin{align}
S_1 & =- {1 \over 2}M^3  N_c^2\, x\, \int d^4 x \, dr  V_f(\l,\t) \sqrt{-\mathrm{det}({\mathcal G})} \biggl( \kappa(\l,\t) \, \tau(r)^2 \, \mathcal{G_S}^{MN} \, (\partial_M \xi +2 A_M) \, (\partial_N \xi +2 A_N)\\ 
&- { w(\l,\t)^2 \over 2} \biggl[ \mgs^{MS} \, A_{ST} \, \mgs^{TN} \, A_{NM} + \mga^{MS} \, A_{ST} \, \mga^{TN} A_{NM}  -  {1 \over 2}\, (\mga^{MN} A_{MN})^2 \biggr] \biggr) \nn \\
 S_2 & = -{1 \over 2} M^3 N_c^2 \, \int d^4 x dr \sqrt{-g}  \, Z(\l) \, \big(\partial_M \a - x(2 V_a \,A_M - \xi \pa_M V_a \big) \big(\partial_N \a - x(2 V_a \,A_N - \xi \pa_N V_a) \big) g^{MN} \nn \,
 \end{align}
 where the "metric" $\mathcal{G}$ was defined in (\ref{DBImet}).
It is noticed that  ${\mathcal G}_{MN}$ is split in its symmetric and antisymmetric parts ${\mathcal G}_{MN}=\mgs_{MN}+\mga_{MN}$.
The fully couple system of the equations of motion in the presence of external magnetic field then reads

\begin{eqnarray}
&&  \pa_M \biggl[ Z(\l) \, \sqrt{-g} \, g^{MN} (\partial_M \a -x(2 V_a \,A_M - \xi \pa_M V_a ) \biggr] =0 \\
&& \partial_M\biggl[ V_f(\l,\t) \, \kappa(\l, \t) \,\tau^2  \, \sqrt{-\mathrm{det}({\mathcal G})} \,{\mathcal G_S}^{MN}\,(\partial_N \xi + 2 A_N) \biggr] \\
&&-  Z(\l) \, \pa_M V_a \, \sqrt{-g} \, g^{MN} \biggl[\partial_M \a + x(2 V_a \,A_M - \xi \pa_M V_a \biggr] =0 \nn \\
&& \partial_N \biggl[ V_f(\l,\t) \, \sqrt{-\mathrm{det}({\mathcal G})} \, w(\l,\t)^2 \biggl( \mgs^{MS} \mgs^{TN} A_{ST} + \mga^{MS} \mga^{TN} A_{ST} \nn\\
&&  + {1\over 2}\, \mga^{ST} \mga^{MN} A_{ST} \biggr) \biggr] -2 \, V_f(\l,\t) \, \sqrt{-\mathrm{det}({\mathcal G})} \,\kappa(\l,\t) \,\tau^2 \,\mgs^{MN}\,(\partial_N \xi + 2 A_N) \nn \\
&&+2 Z(\l) \, V_a \, \sqrt{-g} \, g^{MN} \biggl[\partial_N \a -x(2 V_a \,A_N - \xi \pa_N V_a \biggr]=0 \,.
\end{eqnarray}

\bibliographystyle{JHEP}
\bibliography{Bibliography}

\end{document}